\documentclass[conference]{IEEEtran}
\IEEEoverridecommandlockouts
\usepackage{cite}
\usepackage{amsmath,amssymb,amsfonts}
\usepackage{dsfont}
\usepackage{algorithmic}
\usepackage{graphicx}
\usepackage{textcomp}
\usepackage{xcolor}
\usepackage{hyperref}
\usepackage{todonotes}
\usepackage{algorithm}
\usepackage{tabularx,booktabs}
\usepackage{subcaption}
\usepackage{placeins}

\newcolumntype{Y}{>{\centering\arraybackslash}X}
\usepackage{multirow}
\usepackage[font=small,labelfont=bf]{caption}

\usepackage{subfloat}

\usepackage{amsthm}\theoremstyle{plain}

\def\BibTeX{{\rm B\kern-.05em{\sc i\kern-.025em b}\kern-.08em
    T\kern-.1667em\lower.7ex\hbox{E}\kern-.125emX}}

\begin{document}

\title{The Impact of Adversarial Node Placement in Decentralized Federated Learning Networks
\thanks{This project was supported in part by ONR under grants N000142112472 and N000142212305, and by NSF under grants CNS-2146171 and ITE-2326898}
}

\author{
    \IEEEauthorblockN{Adam Piaseczny\IEEEauthorrefmark{1}, Eric Ruzomberka\IEEEauthorrefmark{2}, Rohit Parasnis\IEEEauthorrefmark{3}, Christopher G. Brinton\IEEEauthorrefmark{1}}
    
    \IEEEauthorblockA{\IEEEauthorrefmark{1}Purdue University, Email: \{apiasecz, cgb\}@purdue.edu}
    
    \IEEEauthorblockA{\IEEEauthorrefmark{2}Princeton University, Email: er6214@princeton.edu}
    
    \IEEEauthorblockA{\IEEEauthorrefmark{3}
    Massachusetts Institute of Technology, Email: rohit100@mit.edu}
    
}
\maketitle

\date{}
\begin{abstract}

As Federated Learning (FL) grows in popularity, new decentralized frameworks are becoming widespread. These frameworks leverage the benefits of decentralized environments to enable fast and energy-efficient inter-device communication. However, this growing popularity also intensifies the need for robust security measures. While existing research has explored various aspects of FL security, the role of adversarial node placement in decentralized networks remains largely unexplored. This paper addresses this gap by analyzing the performance of decentralized FL for various adversarial placement strategies when adversaries can jointly coordinate their placement within a network. We establish two baseline strategies for placing adversarial node: random placement and network centrality-based placement. Building on this foundation, we propose a novel attack algorithm that prioritizes adversarial spread over adversarial centrality by maximizing the average network distance between adversaries. We show that the new attack algorithm significantly impacts key performance metrics such as testing accuracy, outperforming the baseline frameworks by between \(9\%\) and \(66.5\%\) for the considered setups. Our findings provide valuable insights into the vulnerabilities of decentralized FL systems, setting the stage for future research aimed at developing more secure and robust decentralized FL frameworks.

\end{abstract}

\section{Introduction}
\label{sec:introduction}
Federated Learning (FL) \cite{mcmahan2017communication} has emerged as a popular method for distributed machine learning applications across various domains\cite{9084352}. By sharing the workload of the training process across different nodes, FL enables parallel computation and eliminates the need for shared access to training data, enhancing both efficiency and privacy of the training process. While traditional FL frameworks follow the client-server architecture, recent work\cite{9933813, raynal2023decentralized} has focused on the fully decentralized setting in which nodes participating in the learning process exchange their models and conduct distributed consensus via D2D communication. This eliminates the need for a central server and offers advantages in terms of communication efficiency, scalability, and robustness.

However, the decentralized setup also introduces new challenges, particularly in terms of security in the training process. Centralized FL frameworks are known to be vulnerable to attacks on the training process by adversarial clients \cite{lyu2020threats}. These adversaries can employ an array of common and well-established attacks \cite{goodfellow2015explaining, huang2017adversarial, wang2023potent} to compromise the global model's integrity and prevent it from achieving its target. In the absence of a centralized coordinator, we expect that the attack potency is affected not only by the strength of the adversaries, but also by the placement of the adversaries in the network.

In this paper, we investigate this relationship. We aim to answer the following question: \textit{How does the placement of adversarial nodes in the network affect the attack potency?} By addressing this question, we aim to provide valuable insights into the vulnerabilities of decentralized FL systems and lay the groundwork for future research focused on enhancing the security of such frameworks.

\textbf{Related Work}:
While the influence of adversarial placement has not been extensively explored in decentralized FL, it has garnered significant attention in other networked systems \cite{5340022, 7029011}. In many of these systems, the centrality measures of nodes play a pivotal role in understanding the dynamics of worst-case adversarial placement \cite{mirchev2011selective, 10.1145/1811039.1811063}. This insight has been instrumental in formulating strategies for applications like virus defense and detection, underscoring the importance of centrality in network security.

Thus, recent developments in decentralized FL would reflect a growing interest in centrality-based approaches for placing adversarial nodes \cite{yar2023backdoor}. Given this research trend and similarities between decentralized FL and other networked systems, it would be reasonable to assume that centrality is a key factor in determining adversarial impact in FL environments. However, our research presents a nuanced perspective. We demonstrate that centrality-based approaches do not necessarily align with the most effective strategy for adversarial placement, and in fact significantly underperform other placement strategies in certain FL scenarios.

Correspondingly, growing concern over security vulnerabilities in deep learning has led to an increase in research efforts aimed at enhancing the security of FL architectures. A variety of strategies have been proposed\cite{9721118, Cao_Jia_Gong_2021} that modify the aggregation process or fine-tune model parameters for higher training resilience. More recently, some attention has been directed towards the robustness of FL frameworks with different network topologies \cite{raynal2023decentralized}. These studies analyzed various configurations, from semi-centralized to fully decentralized setups. However, many questions remain unanswered, particularly regarding how network topologies and the placement of adversarial nodes interact. This paper aims to address some of these questions and fill this research gap.

\textbf{Summary of Contributions} In this paper, we make the following key contributions:
\begin{itemize}
    \item \textbf{Model Poisoning Attack Analysis in Decentralized FL}: We analyze the effects of adversarial placement in poisoning attacks for different network topologies within decentralized FL settings. Our analysis introduces two baselines for evaluating the potency of placement attack strategies: random choice and centrality-based attacks.
    \item \textbf{MaxSpAN-FL Attack Framework}: We introduce a novel algorithm for placing attack nodes in specific network types. The resulting algorithm, MaxSpAN-FL (Maximally Spread-out Adversaries in a Network), relies on the principle of (approximately) maximizing the average inter-adversarial distance. MaxSpAN-FL demonstrates superior performance over baseline methods, achieving from $9\%$ up to $66.5\%$ performance improvement over next best baseline for the scenarios considered.
\end{itemize}

\section{Methodology}
\label{sec:methodology}
In this section, we first introduce our decentralized FL model in Section \ref{sec:general_setup}. Given this model, we then define our baseline algorithms we used to evaluate adversarial node placement in decentralized FL in Section \ref{sec:baseline_algorithms}, and finally we define our proposed new attack algorithm in Section \ref{sec:new_attack_algorithm}.

\subsection{System Model}\label{sec:general_setup}

\begin{figure}[t]
\centering
\includegraphics[width=0.48\textwidth]{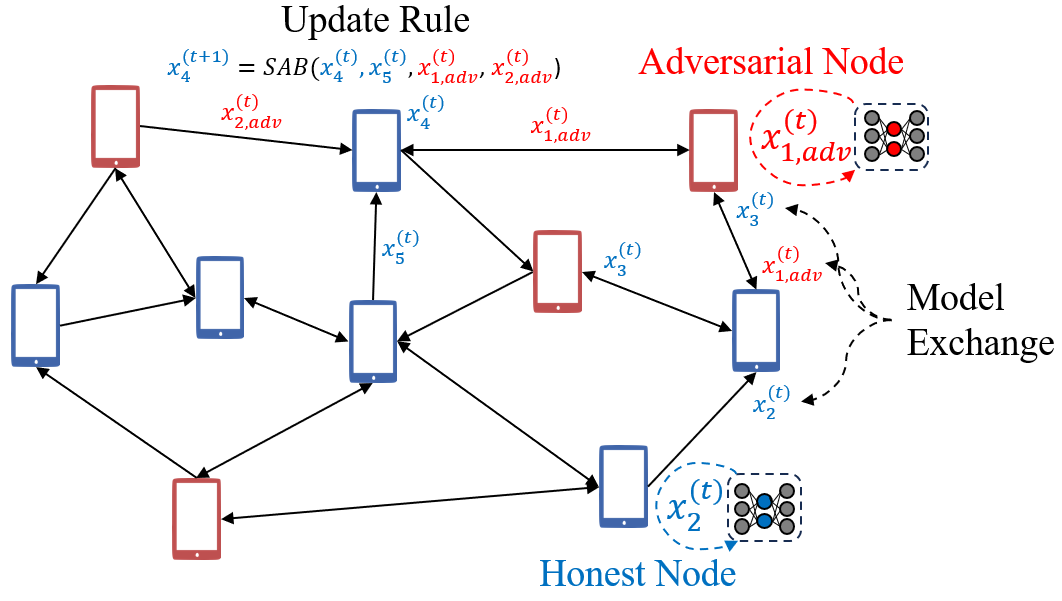}
\caption{Decentralized federated learning with adversarial nodes.}
\label{fig:net_adv}
\vspace{-1em}
\end{figure}

As depicted in Fig. \ref{fig:net_adv}, a network of D2D nodes is a strongly connected, time-invariant, directed graph $G = (V, E)$ where $V$ is the set of nodes and $E$ is the set of edges (i.e., communication links) between the nodes. The set $V$ can be partitioned into two sets: a set of adversarial nodes $A$ and a set of honest nodes $H:=V \setminus A$. 

The network of D2D nodes participates in a decentralized FL scheme over a time horizon $t=1,2,3,\ldots, N_{\text{epochs}}$. At time $t$, an honest node $i \in V\setminus A$ trains a local model $x_i^{(t)} \in \mathbb{R}^p$  using a local data-set $\mathcal{D}_i$ and a local loss function $f_i:\mathbb{R}^p \times \mathbb{R}^{|\mathcal{D}_i|}\rightarrow \mathbb{R}$. Training follows the S-AB aggregation procedure \cite{9029217}, in which the local model is computed as $$x^{(t)}_i = \sum_{j:(j,i)\in E} \frac{x_j^{(t-1)}}{|j:(j,i) \in E|} -\alpha y^{(t-1)}_i$$ where $\alpha>0$ is the learning rate and $y^{(t-1)}_i$ is the global gradient estimation of node $i$ at time $t-1$ as given by the S-AB aggregation procedure. In contrast to the honest nodes, adversarial nodes do not follow the S-AB aggregation procedure. Instead, an adversarial node $i \in A$ mounts an FGSM attack \cite{goodfellow2015explaining} at time $t$ using a local poisoned data-set $\mathcal{D}_{i, adv}^{(t-1)}$ calculated as:$$\mathcal{D}_{i, adv}^{(t-1)} = \mathcal{D}_i + \epsilon_i \cdot \text{sign}(\nabla_{\mathcal{D}_i}f_i(x_i^{(t-1)}, \mathcal{D}_i))$$ where $\epsilon_i > 0$ is the attack power of node $i$. Using the poisoned data, the adversary computes the local model $$x^{(t)}_i = x^{(t-1)}_i - \alpha \nabla_{x}f_i(x^{(t-1)}_i, \mathcal{D}_{i, adv}^{(t-1)})$$ We assume that all adversarial nodes use the same attack power $\epsilon>0$. After training but before time $t+1$, all nodes exchange their local models and global gradient estimations. 

The goal of the honest nodes is to find a global model $x^*$ that minimizes the global loss function
\begin{equation} \label{eq:global_loss}
   f(x) = \sum_{i \in V\setminus A}f_i(x_i, \mathcal{D}_i).
\end{equation}
The S-AB aggregation procedure  guarantees all nodes reach consensus and find an optimal model in the setting where all nodes are honest (i.e., $A$ is empty). The adversaries have a goal opposite of the trusted nodes, namely, to maximize the global loss function (\ref{eq:global_loss}). 

\subsection{Baseline Algorithms}\label{sec:baseline_algorithms}
To evaluate potency of a placement attack on a network we established the following 2 baseline algorithms for placing adversarial nodes: random placement and network centrality-based placement.

The random placement attack serves as our initial and most fundamental baseline for evaluating the effectiveness of more sophisticated attack strategies. In this approach, the subset $A \subset V$ is randomly selected, with the size $|A|$ specified in advance. Given its random nature, the random placement attack does not incorporate any strategic planning or targeted placement, making it the simplest form of attack in our study and providing a basic point of comparison for other attacks. 

The network (eigenvector) centrality-based placement attack serves as our second baseline, offering a more nuanced approach than the random placement attack. This strategy is designed to exploit information that is inherent to the network topology when placing adversaries. It is rooted in the intuitive hypothesis, informed by other fields, that nodes with higher eigenvector centrality could potentially have a greater impact on the network. In this strategy, each node in the network graph $G$ has its eigenvector centrality calculated. For a given size $|A|$, the subset $A$ is formed by selecting the $|A|$ nodes in $V$ with the largest eigenvector centralities. The underlying intuition suggests that these nodes would be most effective in spreading malicious updates across the network.

Eigenvector centrality was chosen for node placement because it is a comprehensive measure that considers the relationship of the node to its neighbors as well as to the entire network. This provides a mathematical basis for analysis, making it an attractive choice for our study. Furthermore, the computational efficiency of calculating eigenvector centrality via power iteration methods adds to its feasibility for larger networks.

Using this algorithm for a baseline will allow us to evaluate the effectiveness of an attack strategy using network topology information in an intuitive way and quantify the potency and success of other more sophisticated attacks. 

\subsection{New Attack Algorithm}\label{sec:new_attack_algorithm}

In this section we propose an adversarial placement attack termed the Maximally Spread-out Adversaries in a Network for Federated Learning (MaxSpAN-FL) placement attack. The MaxSpAN-FL placement attack is designed to strategically identify influence regions for adversarial nodes in a way that efficiently and approximately maximizes the \textit{average distance between adversarial nodes} $$d_{\text{avg}} = \frac{1}{{|A| \choose 2}}\sum_{i \in A} \sum_{\substack{j \in A \\ j>i}}d(i,j)$$ where $d(i,j)$ is the shortest graph hop distance between nodes $i$ and $j$ on $G$. It does so by employing a breadth-first search (BFS) clustering technique to consider the structure of the neighborhood surrounding a node. This allows us to spread the adversaries throughout the network in an efficient way, which we hypothesize should lead to a stronger attack.

The algorithm is outlined in \ref{alg_MaxSpAN-FL} and employs a breadth-first search (BFS) clustering technique to identify influence regions for each node in the network. Initially, the algorithm calculates these influence regions for all nodes \( g \in G \) using BFS. The BFS is run until the number of nodes in a given cluster exceeds the pre-determined cluster area \( S_{\text{cluster}} \). This area is calculated by dividing the total number of nodes \( n_{\text{clients}} \) by the number of adversaries \( n_{\text{advs}} \), ensuring that each adversary has a roughly equal area of influence.

Once these influence regions are established, the algorithm enters an iterative phase where it selects adversarial nodes \( a_{\text{best}} \) in a way that minimizes the overlap between their influence regions. Specifically, for each node still considered `honest', the algorithm calculates the overlap \( o \) of its influence region with those of the already selected adversarial nodes. This overlap is quantified as 
\begin{equation}
    o = \left|\mathcal{C}_G[g] \cap \left( \bigcup_{a \in A} \mathcal{C}_G[a] \right) \right|
\end{equation}
The node with the least overlap is then selected as the next adversarial node, added to the set \( A \) of adversarial nodes, and removed from the set \( H \) of honest nodes.

Notice that the algorithm introduces a degree of randomness in the placement of adversarial nodes. The first node is chosen completely at random, providing a starting point for the algorithm. Subsequent nodes are selected based on minimizing overlap, but within that constraint, the placement is influenced by the current state of the network and the nodes already chosen as adversaries.

\begin{algorithm}
\caption{MaxSpAN-FL Attack Algorithm}

\begin{algorithmic}[1]
\STATE \textbf{Input:} Graph \( G \), Number of adversaries \( n_{\text{advs}} \), Number of clients \( n_{\text{clients}} \)
\STATE \textbf{Output:} List of adversarial nodes \( A \)
\STATE \( \text{Cluster Area } S_{\text{cluster}} \leftarrow \lfloor n_{\text{clients}} / n_{\text{advs}} \rfloor \)
\STATE Cluster dictionary \(\mathcal{C}_G \leftarrow \{g:\{\}\}_{g \in G}\) 
\FOR{\(g \in G \)}
    \STATE \( \mathcal{C}_G[g] \leftarrow BFS\_Cluster(G, g, S_{\text{cluster}})\)
\ENDFOR
\STATE Set of adversaries \(A \leftarrow \{\}\)
\STATE Set of honest nodes \(H \leftarrow G\)
\WHILE{\( |A| < n_{\text{advs}} \)}
    \STATE \( o_{\min} \leftarrow \infty \)
    \STATE \( a_{\text{best}} \leftarrow \text{None} \)
    \FOR{\(g \in H \)}
        \STATE \(o = \left|\mathcal{C}_G[g] \cap \left(\underset{a \in A}{\cup}\mathcal{C}_G[a]\right)\right|\)
        \IF{\(o < o_{\min}\)}
            \STATE \( o_{\min} \leftarrow o\)
            \STATE \( a_{\text{best}}\leftarrow g \)
        \ENDIF
    \ENDFOR
    \STATE \( A \leftarrow A \cup {a_{\text{best}}}\)
    \STATE \( H \leftarrow H \setminus {a_{\text{best}}}\)
\ENDWHILE
\RETURN \( A \)
\end{algorithmic}

\end{algorithm}\label{alg_MaxSpAN-FL}

\section{Results and Discussion}
\label{sec:results_and_disc}
Here, we first discuss the experimental setup in Section~\ref{exp_setup}, which includes the network architectures, model architectures, and datasets used for evaluation. Then, in Section~\ref{exp_results}, we present the experiments conducted and discuss the corresponding results for various hyperparameter settings.
\subsection{Experimental Setup}\label{exp_setup}
In our experiments, we consider the decentralized FL model of Section \ref{sec:general_setup}. We evaluate this model for distinct network topologies in which the graph $G$ is drawn at random. We consider two random graph distributions: Directed Geometric (DG) graphs and Erdos-Renyi (ER) graphs. To assess the effectiveness of our proposed MaxSpAN-FL Attack we synthesize networks with varying hyperparameters, such as network connectivity or network size. For our experiments we generate between 20 and 50 network realizations, using different random seeds.

\begin{figure*}[t]
  \centering
\includegraphics[scale=0.43]{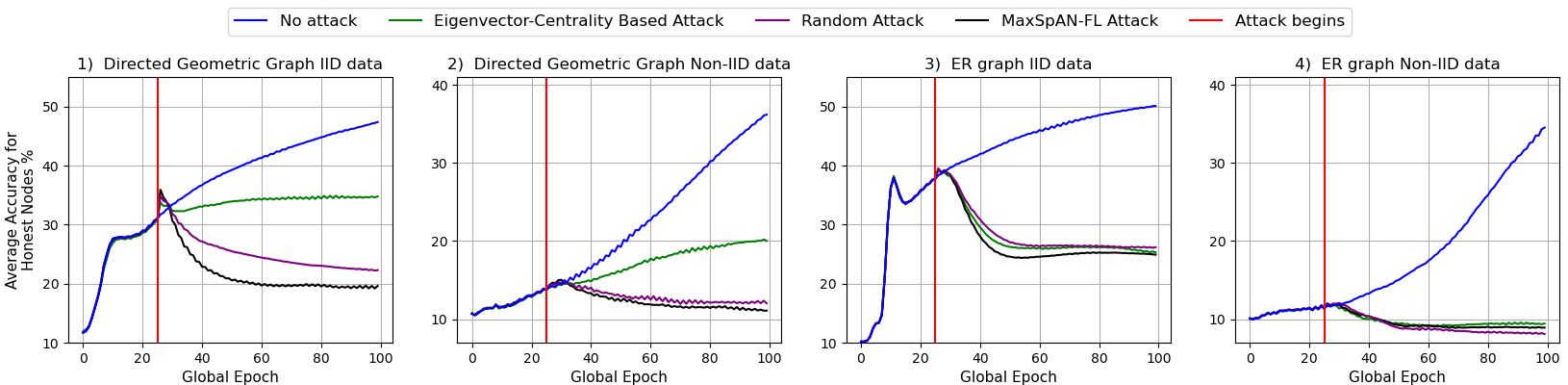}
  \caption{Average testing accuracy of honest nodes in 25-node networks, comparing Directed Geometric graphs with connection radius \( r = 0.2 \) and Erdős–Rényi graphs with edge probability \( p = 0.5 \), for both IID and Non-IID data distributions. The effects of various attack placement strategies on the network's performance are illustrated. Adversarial percentage is $20\%$.}
  \label{fig:main_result}
  \vspace{-1.3em}
\end{figure*}

The network hyperparameters we consider are as follows:
\begin{itemize}
    \item Connectivity parameters for 25-node networks for both graph types. For the DG graphs, we randomly assign nodes' locations in a 2D unit square and vary the connection radius parameter $r$ to $0.2$, $0.4$, and $0.6$ to simulate different levels of network density. Correspondingly, for ER graphs, we adjust the edge creation probability $p$, setting it to $0.1$, $0.3$, and $0.5$. 
    \item Size parameters for the DG graph with $r = 0.2$. We consider network sizes of 10, 25, 50, and 100, and for each network we consider different adversarial percentages, ranging from $5\%$ up to $20\%$.
\end{itemize}

Following network generation, each node trains an image classification model. We use the Fashion-MNIST (FMNIST) dataset for evaluation, distributing it across nodes in both IID and Non-IID manners, with the latter having 3 classes per node. Tailored Convolutional Neural Network (CNN) model architecture serves as our classifier. The model architecture consists of two convolutional layers, each succeeded by a dropout layer, a ReLU activation function, and a max-pooling operation, culminating in two dense layers for classification. 

Each node executes a predefined number of local training iterations on their respective model. Upon completion, the models are distributed amongst neighboring nodes via the previously mentioned S-AB algorithm. We use Adam optimizer for training.

In all cases, we evaluate the 2 baseline placement attacks, as well as the MaxSpAN-FL attack for selecting the group of adversarial nodes that run FGSM as described in \ref{sec:general_setup}. Adversarial nodes always ignore models from their in-neighbors.\footnote{The source code supporting our experiments is publicly available at \url{https://github.com/Adampi210/MaxSpANFL_atck_code_data.git}.}

\subsection{Experimental Results}\label{exp_results}
\subsubsection{Node Placement}
Figure \ref{fig:main_result} illustrates that the adversarial placement attack potency depends on both the network's graph structure and the data distribution among clients. In directed geometric (DG) graphs with IID data, our proposed MaxSpAN-FL attack method surpasses alternative strategies, achieving a minimal final test accuracy of \(19.6\%\), outperforming the \(22.3\%\) and \(34.8\%\) final accuracies achieved by random node and eigenvector-centrality based attacks, respectively. However, the performance disparity narrows in other scenarios, with the MaxSpAN-FL attack outperforming the next best method by only \(0.93\%\) in DG graphs with Non-IID data and \(0.63\%\) in ER graphs with IID data. In ER graphs with Non-IID data, our method trails slightly behind the most effective attack, with a marginal performance gap of \(0.82\%\).

Intuitively, the similar performance observed across placement attack strategies in ER graphs follows from the high degree of randomness in ER: with large probability, all nodes in an ER graph have similar centrality scores and any two nodes have a small shortest path length. Thus, our three placement attack strategies are nearly equivalent in ER graphs.

Conversely, in DG graphs with IID data, our MaxSpAN-FL algorithm not only demonstrates superior final performance but also exhibits a faster rate of decreasing test accuracy. The algorithm reaches its lowest accuracy point around the 60th global epoch, a point not achieved by other methods even by the 100th epoch. This highlights the significance of strategic adversarial placement in DG graphs.

Lastly, the data distribution among clients significantly influences the performance gap between placement attack strategies. In Non-IID scenarios, this gap notably reduces, potentially due to the added complexity of class distribution in adversarial node placement and the varying degrees of model convergence at the time of attack. Notably, the test accuracy of attacked clients in Non-IID settings is substantially lower than in IID scenarios, further underscoring the role of data distribution in shaping attack outcomes.

\subsubsection{Effects of Connectivity}
\begin{figure}[t]
\centering

\begin{subfigure}{\columnwidth}
  \centering
  \includegraphics[width=\linewidth]{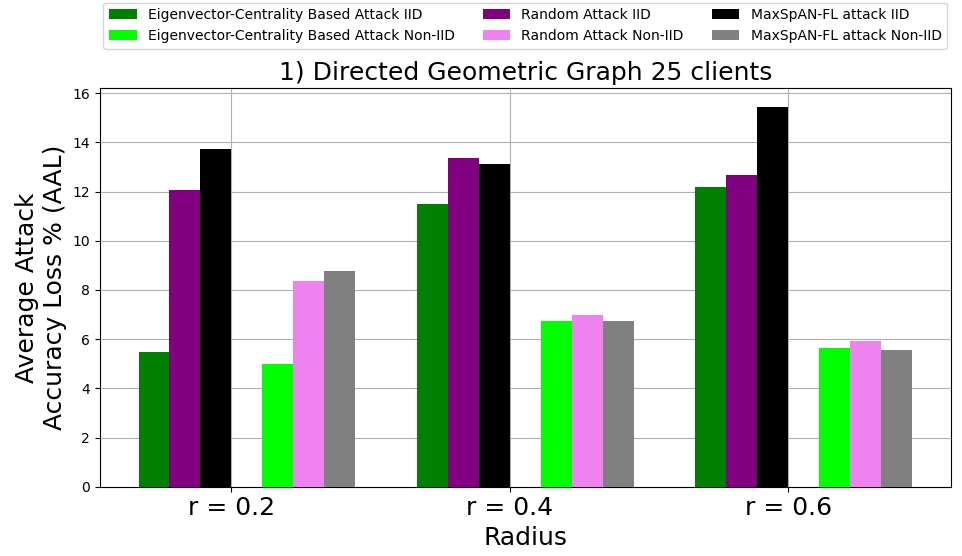}
\end{subfigure}%
\vspace{-0.15em}
\begin{subfigure}{\columnwidth}
  \centering
  \includegraphics[width=\linewidth]{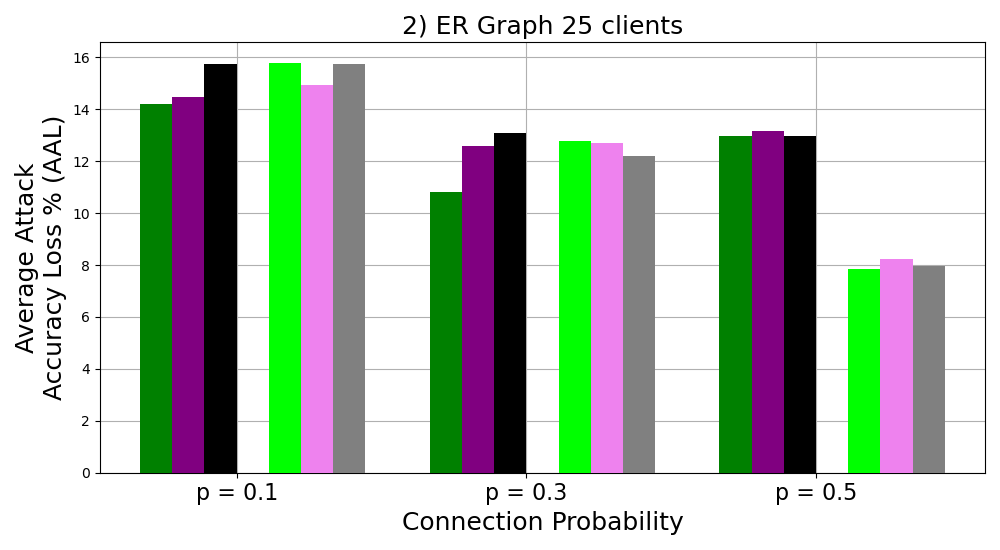}
\end{subfigure}
\vspace{-1.5em}
\caption{Average Attack Accuracy Loss for Directed Geometric and ER graphs with 25 nodes and $20\%$ adversaries for different connectivity parameters.}
\label{fig:connectivity_results}
\vspace{-1.5em}
\end{figure}

Figure \ref{fig:connectivity_results} demonstrates the influence of network connectivity on the potency of different placement attacks, considering both IID and Non-IID data distributions in networks with 25 nodes. The analysis of this setting is based on the Attack Accuracy Loss ($\text{AAL}$) measure, calculated as $$\text{AAL} = \overset{N_{\text{epochs}}}{\underset{i = t_{\text{attack}}}{\sum}}\left[\text{(\% acc no attack)[i]} - \text{(\% acc with attack)[i]}\right]$$
as it measures both by how much and how fast the attack decreases the testing accuracy. Higher AAL indicates better attack potency.
Attack performance advantage is computed as: $$\frac{\text{AAL}_{\text{best attack}}-\text{AAL}_{\text{next best attack}}}{\text{AAL}_{\text{next best attack}}} \times 100\%$$

In Directed Geometric (DG) graphs, connectivity is varied with radius values \( r \in \{0.2, 0.4, 0.6\} \). Our MaxSpAN-FL attack demonstrates superior performance in the \( r = 0.2 \) and \( r = 0.6 \) scenarios under IID conditions, outperforming the next best attack by \( 14.0\% \) and \( 21.8\% \), respectively. However, in the \( r = 0.4 \) scenario, the performance margin narrows, with our attack trailing by a a slight margin of \( 1.8\% \). These findings indicate that MaxSpAN-FL maintains a competitive edge across varying connectivity levels in DG graphs.

For Erdős-Rényi (ER) graphs, we assess the impact of different connection probabilities \( p \in \{0.1, 0.3, 0.5\} \). The MaxSpAN-FL attack shows its greatest advantage in the \( p = 0.1 \) case, surpassing the next best attack by approximately \( 9\% \). As the connection probability increases, the performance gap narrows, aligning with intuition that more sparse ER graphs exhibit greater variance in edge distribution, making network data more informative.

These results not only affirm the performance edge and competitiveness of the MaxSpAN-FL attack across diverse network configurations but also offer several intriguing insights. Notably, in Non-IID scenarios, the peak Attack Accuracy Loss consistently decreases with increased graph connectivity, a trend not as pronounced in IID settings. This is likely caused by enhanced information exchange among honest clients in denser networks. In ER graphs, as connectivity approaches that of a fully connected network, the placement attack performance differences go to 0, reinforcing the notion that node placement becomes less critical in highly connected networks. Additionally, the Eigenvector-Centrality Based attack improves its performance by $123.4\%$ in DG graphs with IID data as connectivity increases, suggesting better connectivity to most central nodes, even thought it still underperforms MaxSpAN-FL for all radius values.

\subsubsection{Effects of Network Size and Adversarial Percentage}
\begin{figure}[t]
\centering
\begin{subfigure}{\columnwidth}
  \centering
  \includegraphics[width=\linewidth]{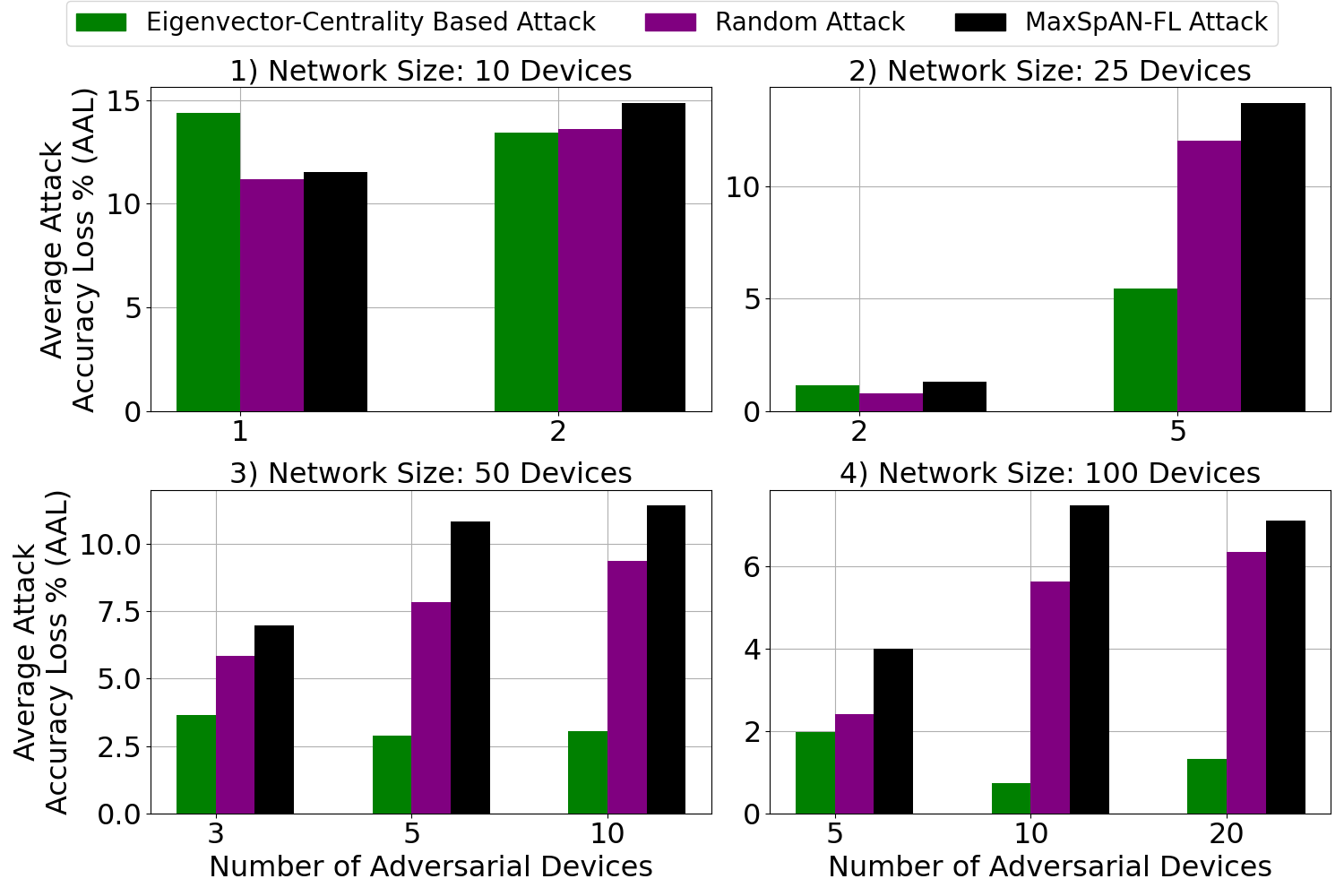}
\end{subfigure}%
\vspace{-0.25em}
\caption{Average Attack Accuracy Loss for Directed Geometric graphs with $r = 0.2$ and IID data distribution, for different network sizes and number of adversaries}
\label{fig:network_sizes}
\vspace{-1em}
\end{figure}

The impact of network size and adversarial percentage on the performance differential among various attack strategies is illustrated in Figure \ref{fig:network_sizes}. 

We consider the Directed Geometric (DG) graphs with radius \( r = 0.2 \), since ER graphs inherently are generated using IID process, therefore their performance across various sizes is consistent. Additionally, we focus on IID data distribution, as Non-IID scenarios did not yield significant variances in attack performance to discern clear trends.

In nearly all examined cases, the MaxSpAN-FL attack surpasses other methods, particularly notably at adversarial percentages of \( 10\% \) and \( 20\% \) in larger networks. The attack achieves performance advantage of up to \( 66.5\% \) improvement over the closest competing attack in a 100-node network with 5 adversaries. Even in the smallest network scenario of 10 clients and 2 adversaries, MaxSpAN-FL maintains a \( 9.4\% \) advantage, which is it's lowest perforance difference that's non-negative. The sole exception occurs in a 10-client network with a single adversary, where the Eigenvector-Centrality based attack excels, outperforming MaxSpAN-FL by approximately \( 20.1\% \). This is an expected outcome, as MaxSpAN-FL's initial node placement is random, equating to a random node attack in single-adversary scenarios. Conversely, the Eigenvector-Centrality approach targets the most influential node, thereby maximizing attack potency. This exception aside, MaxSpAN-FL consistently outperforms other strategies across varying network sizes in DG graphs, underscoring the effectiveness of the proposed placement attack method.

\subsubsection{Effects of Attack Deployment Time}
\begin{figure}[ht]
\centering

\begin{subfigure}{\columnwidth}
  \centering
  \includegraphics[width=\linewidth]{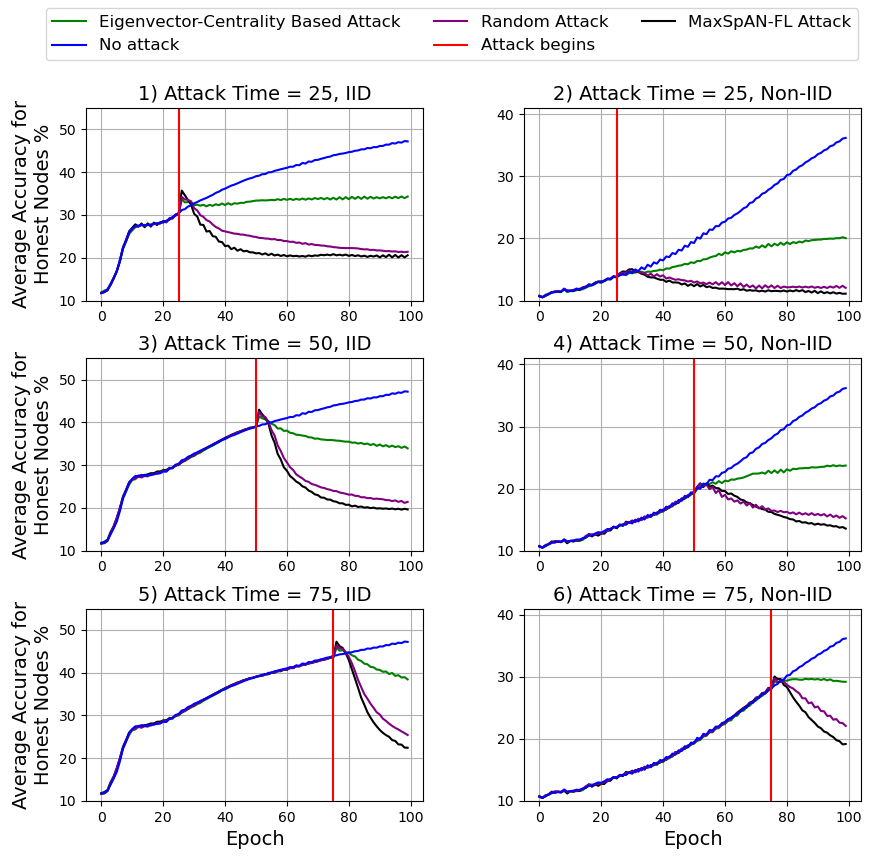}
\end{subfigure}%
\vspace{-0.25em}
\caption{Average testing accuracy over honest nodes in 25-node Directed Geometric Graph with $r=0.2$, for both IID and Non-IID data distributions for different attack deployment times. Adversarial percentage is $20\%$.}
\label{fig:timing_dir_geom}
\vspace{-1em}
\end{figure}

Our final experiments investigate the impact of varying attack deployment times during global aggregation. We continue to focus on the Directed Geometric (DG) graph with 25 clients and radius \( r = 0.2 \), for previously stated reasons. The findings are depicted in Figure \ref{fig:timing_dir_geom}.

Across various attack timings, the MaxSpAN-FL attack consistently demonstrates superior performance in terms of both final model accuracy and adversarial accuracy loss. This suggests that the effectiveness of the proposed attack is largely unaffected by its deployment time. Notably, in non-IID scenarios, the  final accuracy discrepancy widens with later attack times. This trend could intuitively be attributed to the degree of model
convergence at the point of attack, although this requires further investigation. Additionally, while attacks initiated at later stages do not reach convergence within 100 epochs, their trajectories seem to be aligning towards final accuracy levels similar to those observed in the 25-epoch attack scenario.

\subsection{Key Takeaways}
\begin{itemize}
    \item The impact of adversarial node placement is significantly influenced by the network's graph structure and the data distribution, as shown in Figure \ref{fig:main_result}.
    \item In most scenarios, MaxSpAN-FL either surpasses or matches baseline approach performances. MaxSpAN-FL performance improvement reaches up to around $66.5\%$.
    \item Network connectivity plays an important role in determining the effectiveness of different adversary placement strategies and their relative performance. 
    \item Varying graph sizes maintain the performance disparity between adversarial placement strategies. The proportion of adversarial nodes notably influences this difference, particularly in scenarios with fewer adversarial nodes.
    \item The timing of the attack consistently maintains performance difference trend between different placement strategies. Networks with extended training rounds exhibit larger variations in the degree of performance disparity among these strategies. 
\end{itemize}

\section{Conclusion and Future Work}
\label{sec:conclusion}
The rapid advancement of decentralized processing frameworks introduces a range of novel applications, while simultaneously presenting unique security challenges. In this paper we delved into the impact of adversarial node placement within FL environments, a critical aspect often influenced by network topological factors. Our comprehensive analysis reveals that the effect of adversarial node placement on attack potency in FL depends on a variety of elements, including network type, data distribution, connectivity, size, and the timing of the attack. We also developed a novel attack placement algorithm, and demonstrated superior performance in attack potency across diverse conditions. The attack performance suggests that under various network conditions, there is a combination of nodes that performs better on average. In future work we plan to explore additional graph types, centrality measures, and the role of data heterogeneity in attack strategies.


\bibliography{ref.bib}
\bibliographystyle{IEEEtran}

\end{document}